\documentclass[                                                                                                           
 reprint,
 amssymb, amsmath,
 aip,JCP
]{revtex4-1}

\usepackage{bm} 
\usepackage{morefloats}
\usepackage{comment} 
\usepackage{dcolumn} 
\usepackage[colorlinks=true,linkcolor=blue]{hyperref}%
\usepackage[table]{xcolor} 
\usepackage{array} 
\expandafter\ifx\csname package@font\endcsname\relax\else
 \expandafter\expandafter
 \expandafter\usepackage
 \expandafter\expandafter
 \expandafter{\csname package@font\endcsname}%
\fi
\usepackage{xspace}
\usepackage{soul}
\usepackage{graphicx}
\usepackage{subcaption}
\usepackage{pgfplots}
\usepackage{pgfplotstable}
\usepackage{rotating} 
\usepackage{amsmath}

\usepackage[normalem]{ulem}
\usepackage[utf8]{inputenc}

\newcommand{\deriv}[3]{\frac{\text{d}^{#3}#1}{\text{d}#2^{#3}}\bigg|_{#2 = 0}}

\newcommand{\ket}[1]{{\ensuremath{|#1\rangle}\xspace}}
\newcommand{\bra}[1]{{\ensuremath{\langle #1|}\xspace}}

\newcommand{\basis}[0]{\mathcal{B}}

\newcommand{\br}[1]{{\mathbf{r}_{#1}}}

\newcommand{\brb}[1]{{\bf r}_{#1}}
\newcommand{\rab}[1]{|\brb{1} - \brb{2}|}

\newcommand{\wbasiscoal}[1]{W^{\basis}(\br{})}

\begin{document}

\author{Diata Traore}
\email{diata.traore@upmc.fr}
\affiliation{Laboratoire de Chimie Théorique, Sorbonne Université and CNRS, F-75005 Paris, France}
\author{Julien Toulouse}
\email{toulouse@lct.jussieu.fr}
\affiliation{Laboratoire de Chimie Théorique, Sorbonne Université and CNRS, F-75005 Paris, France}
\affiliation{Institut Universitaire de France, F-75005 Paris, France}
\author{Emmanuel Giner}%
\email{emmanuel.giner@lct.jussieu.fr}
\affiliation{Laboratoire de Chimie Théorique, Sorbonne Université and CNRS, F-75005 Paris, France}

\title{Basis-set correction for coupled-cluster estimation of dipole moments}

\date{February 9, 2022}

\begin{abstract}
The present work proposes an approach to obtain a basis-set correction based on density-functional theory (DFT) for the computation of molecular properties in wave-function theory (WFT). This approach allows one to accelerate the basis-set convergence of any energy derivative of a non-variational WFT method, generalizing previous works on the DFT-based basis-set correction where either only ground-state energies could be computed with non-variational wave functions \href{https://pubs.acs.org/doi/10.1021/acs.jpclett.9b01176}{[J. Phys. Chem. Lett. \textbf{10}, 2931 (2019)]} or properties where computed as expectation values over variational wave functions \href{https://aip.scitation.org/doi/10.1063/5.0057957}{[J. Chem. Phys. \textbf{155}, 044109 (2021)]}. 
This work focuses on the basis-set correction of dipole moments in coupled-cluster with single, double, and perturbative triple excitations (CCSD(T)), which is numerically tested on a set of fourteen molecules with dipole moments covering two orders of magnitude. As the basis-set correction relies only on Hartree-Fock densities, its computational cost is marginal with respect to the one of the CCSD(T) calculations. Statistical analysis of the numerical results shows a clear improvement of the basis convergence of the dipole moment with respect to the usual CCSD(T) calculations. 
\end{abstract}

\maketitle

\section{Introduction}\label{sec:Intro}
Quantum chemistry aims to provide theoretical methods to predict molecular properties starting from the many-body quantum mechanical problem. To solve this problem a wide range of methods were developed in the last decades mainly based on wave-function theory (WFT) and density-functional theory (DFT). The purpose of both approaches is to accurately treat correlation effects, or in other terms, the quantum effects which go beyond a mean-field description such as Hartree-Fock (HF).
In the context of WFT, there exists a wide range of methods of increasing computational cost -- ranging from M{\o}ller-Plesset perturbation theory\cite{mp} to coupled-cluster methods\cite{review_cc_bartlett} -- which in principle systematically converge to the full configuration interaction (FCI) limit which is the exact solution within a given basis set. 
Nevertheless, the accuracy of the results of a WFT method -- even at the FCI level -- strongly depends on the quality of the basis set, mainly because of the slow convergence of the wave function near the electron-electron coalescence point\cite{Hyl-ZP-29,Kat-CPAM-57}. 
The combination of the slow basis-set convergence and the strong increase of the computational cost with both the size of the basis set and the number of electrons makes it very difficult to obtain well converged WFT calculations on large systems.

There are mainly two approaches to tackle the basis-set convergence problem of WFT: basis-set extrapolation techniques~\cite{HelKloKocNog-JCP-97,HalHelJorKloKocOlsWil-CPL-98} and explicitly correlated F12 methods\cite{Ten-TCA-12,TenNog-WIREs-12,HatKloKohTew-CR-12, KonBisVal-CR-12, GruHirOhnTen-JCP-17, MaWer-WIREs-18}. 
The basis-set extrapolation techniques rely on a known asymptotic behavior of the correlation energy with the 
size of the basis set but requires WFT calculations with basis sets of increasing sizes, which makes their application limited to small or medium system size. 
The F12 methods accelerate the basis-set convergence of the results thanks to the inclusion of a correlation factor 
explicitly depending on electron-electron distances and restoring Kato's electron-electron cusp condition\cite{Kat-CPAM-57}. Although F12 methods improve indeed the results (typically, energy differences obtained with a F12 method using a triple-zeta basis set are as accurate as the ones obtained with the corresponding uncorrected WFT method using a quintuple-zeta basis set\cite{TewKloNeiHat-PCCP-07}),
the F12 methods necessarily induce computational overheads due to the large auxiliary basis sets required to resolve three- and four-electron integrals\cite{BarLoo-JCP-17}.

An alternative path has been recently introduced by some of the present authors in Ref.~\onlinecite{GinPraFerAssSavTou-JCP-18} where a rigorous framework was proposed to correct for the basis-set incompleteness of WFT using DFT. A central idea of this work is the fact that the Coulomb electron-electron interaction projected in an incomplete basis set is non-divergent and quite similar to the long-range interaction used in range-separated DFT (RSDFT). A basis-set correction density functional can then be built from RSDFT short-range correlation functionals using a local range-separation parameter which automatically adapts to the basis set used. This results in a relatively cheap way of correcting the basis-set incompleteness of WFT, which has the desirable property of leading to an unaltered complete-basis-set (CBS) limit.
Two versions of this theory were proposed: 
(i) a non self-consistent version where the basis-set correction functional is evaluated with any accurate approximation of the FCI density and then simply added to an approximation of the FCI energy in a given basis set\cite{GinPraFerAssSavTou-JCP-18}; 
and (ii) a recently introduced self-consistent variant were the energy is minimized in the presence of 
the basis-set correction functional and therefore allows for the wave function to be changed by the DFT correction\cite{GinTraPraTou-JCP-21}. 
The efficiency of the non-self-consistent approach for computing total energies and chemically relevant energy differences of relatively large magnitudes (such as ionization potentials\cite{GinPraFerAssSavTou-JCP-18,LooPraSceGinTou-JCTC-20}, 
molecular atomization energies\cite{LooPraSceTouGin-JPCL-19,GinSceLooTou-JCP-20,YaoGinLiTouUmr-JCP-21,YaoGinAndTouUmr-JCP-21}, or excitation energies\cite{GinSceTouLoo-JCP-19}) has been well established in previous works on a quite wide range of atomic and molecular systems including light to transition-metal elements, and it was numerically shown that the self-consistent framework does not give any significant improvement of total energies\cite{GinTraPraTou-JCP-21}.  

The main advantage of the self-consistent formulation is nevertheless to allow for the computation of first-order properties as expectation values over the minimized wave function thanks to the variational property of the theory.
In Ref.~\onlinecite{GinTraPraTou-JCP-21} the present authors have focussed on the dipole moments which are known  
to exhibit a slow convergence with respect to the size of the basis set\cite{HalKloHelJor-JCP-99,BakGauHelJorOls-CPL-00,HaiHea-JCTC-18}. 
It was shown that the dipole moments computed at near FCI level with the self-consistent basis-set correction method  
are close to the CBS limit in triple-zeta basis sets, which contrasts with the slow basis-set convergence of the usual WFT approaches. 
The drawback of the self-consistent framework is nonetheless to require a self-consistent variational WFT calculation, which therefore excludes its application to non-variational approaches such as coupled-cluster with singles, doubles, and perturbative triple excitations (CCSD(T)). 

In the present work, we propose to overcome this limitation and target the computation of first-order molecular properties as energy derivatives of the non-self-consistent basis-set correction approach. 
We apply this strategy to the computation of dipole moments at the CCSD(T) level 
and propose a cheap computational strategy for the basis-set correction which uses only densities at the HF level, similarly to what have been done in the context of atomization energies in Ref.~\onlinecite{LooPraSceTouGin-JPCL-19}. 

The paper is organized as follows. In Sec. \ref{sec:Theory}, we introduce the theory of the basis-set correction extended to the computation of dipole moments. 
In Sec. \ref{sec:Comput}, we provide computational details of our study on a set of fourteen molecules with dipole moments covering two orders of magnitude. The numerical results are discussed in Sec. \ref{sec:Results}, and compared for some molecules with the fully self-consistent formalism of Ref.~\onlinecite{GinTraPraTou-JCP-21}. 
Detailed results, as well as the molecular geometries used, are available in the Supplementary material.

\section{Theory}
\label{sec:Theory}

\subsection{Dipole moment from the self-consistent basis-set correction}

In this section, we generalize the framework of the basis-set correction to the presence of a static external electric field.   
Consider the Hamiltonian of a $N$-electron system under an external electric field $\bm{\epsilon}=\epsilon \bm{u}$ of strength $\epsilon$ along a direction $\bm{u}$,
\begin{equation}
	\label{eq:Hamiltonian}
	\hat{H} (\epsilon) = \hat{H}_0 - \epsilon \hat{d},
\end{equation}
where $\hat{H}_0$ is the Hamiltonian of the system without the electric field,
\begin{equation}
	\label{eq:HamiltonianZero}
	\hat{H}_0 = \hat{T} + \hat{V}_{\text{ne}} + \hat{W}_{\text{ee}},
\end{equation}
with the kinetic-energy operator $\hat{T}$, the electron-nuclei interaction operator $\hat{V}_{\text{ne}}$, and the electron-electron interaction operator $\hat{W}_{\text{ee}}$, and $\hat{d}=\hat{\bm{d}} \cdot \bm{u}$ where $\hat{\bm{d}}$ is the total (electron+nuclear) dipole-moment operator,
\begin{equation}
	\label{eq:dipole_total}
		\hat{\bm{d}} = -\sum_{i=1}^{N} \bm{r}_{i} + \sum_{A=1}^{N_\text{nuclei}} Z_A \bm{R}_A, 
\end{equation}
where $\bm{r}_{i}$ are the electron coordinates, and $Z_A$ and $\bm{R}_A$ are the nuclei charges and coordinates.

In the basis-set correction formalism~\cite{GinPraFerAssSavTou-JCP-18,GinSceLooTou-JCP-20,GinTraPraTou-JCP-21}, the ground-state energy $E_0(\epsilon)$ of the Hamiltonian in Eq. \eqref{eq:Hamiltonian} is approximated by
\begin{equation}
\label{eq:basissetcorr}
 \begin{aligned}
		E_{0}^{\mathcal{B}} (\epsilon) 
           & = \min_{\Psi^{\mathcal{B}}} \Bigg\{ \bra{\Psi^{\mathcal{B}}} \hat{H}(\epsilon) \ket{\Psi^{\mathcal{B}}} + \bar{E}^{\mathcal{B}}[n_{\Psi^{\mathcal{B}}}] \Bigg\},
 \end{aligned}
\end{equation}
where the minimization is performed over the set of $N$-electron wave functions $\Psi^{\mathcal{B}}$ expanded on the $N$-electron Hilbert space generated by the one-electron basis set $\mathcal{B}$ and $\bar{E}^{\mathcal{B}}[n_{\Psi^{\mathcal{B}}}]$ is the basis-set correction functional evaluated at the density $n_{\Psi^{\mathcal{B}}}$ of $\Psi^{\mathcal{B}}$. The energy functional $\bar{E}^{\mathcal{B}}[n]$ (introduced in Ref.~\onlinecite{GinPraFerAssSavTou-JCP-18}) compensates for the restriction on the wave functions $\Psi^{\mathcal{B}}$ due to the incompleteness of the basis set $\basis$. The restriction coming from the basis set $\basis$ in Eq.~\eqref{eq:basissetcorr} then applies only to densities $n_{\Psi^{\mathcal{B}}}$. Roughly speaking, since the density converges much faster than the wave function with respect to the basis set, $E_{0}^{\mathcal{B}} (\epsilon)$ is a much better approximation to the exact energy $E_0(\epsilon)$ than the corresponding FCI ground-state energy $E_{\text{FCI}}^{\mathcal{B}} (\epsilon)$ calculated with the same basis set $\basis$. Moreover, in the CBS limit, $\bar{E}^{\mathcal{B}}[n]$ vanishes and thus $E_{0}^{\mathcal{B}} (\epsilon)$ correctly converges to the exact energy $E_{0}(\epsilon)$.

From the basis-set corrected energy $E_{0}^{\mathcal{B}} (\epsilon)$ in Eq.~\eqref{eq:basissetcorr}, one can 
define the corresponding basis-set corrected dipole moment $d^{\mathcal{B}}$ as the first-order derivative with respect to the electric field
\begin{equation}
 \label{eq:def_dipole_b}
 d^{\mathcal{B}} = - \frac{\text{d} E_{0}^{\mathcal{B}} (\epsilon)}{\text{d} \epsilon} \bigg|_{\epsilon=0}.
\end{equation}
It is important to stress here that $d^{\mathcal{B}}$ is different from the FCI dipole moment $d^{\mathcal{B}}_{\text{FCI}}$ with the same basis set $\basis$, as the former is taken as the derivative of $E_{0}^{\mathcal{B}} (\epsilon)$ which contains the basis-set correction functional $\bar{E}^{\mathcal{B}}[n]$. Similarly to the case of the energy, we expect $d^{\mathcal{B}}$ to have a faster basis-set convergence than $d^{\mathcal{B}}_{\text{FCI}}$.

Since $E_{0}^{\mathcal{B}} (\epsilon)$ is stationary with respect to $\Psi^{\mathcal{B}}$, the Hellmann-Feynman theorem applies and gives $d^{\mathcal{B}}$ as a simple expectation value
\begin{equation}
	\label{eq:SCFDipole}
	d^{\mathcal{B}} = \bra{\Psi_{0}^{\mathcal{B}}(\epsilon=0)} \hat{d} \ket{\Psi_{0}^{\mathcal{B}}(\epsilon=0)},
\end{equation}
where $\Psi_{0}^{\mathcal{B}}(\epsilon=0)$ is the minimizing wave function of the self-consistent equation in Eq.~\eqref{eq:basissetcorr} at $\epsilon = 0$. This was the approach used in Ref.~\onlinecite{GinTraPraTou-JCP-21}.

\subsection{Dipole moment from the non-self-consistent basis-set correction}

As initially proposed in Ref.~\onlinecite{GinPraFerAssSavTou-JCP-18} for the case without the electric field, one can avoid the minimization in Eq.~\eqref{eq:basissetcorr} and approximate the energy $E_{0}^{\mathcal{B}}(\epsilon)$ by approximating the minimizing wave function $\Psi_{0}^{\mathcal{B}}(\epsilon)$ in Eq.~\eqref{eq:basissetcorr} by the FCI wave function $\Psi^{\mathcal{B}}_{\text{FCI}}(\epsilon)$ in a given basis set $\basis$. This leads to the following estimation of the ground-state energy 
\begin{equation}
 \label{e_b_approx}
 E_{0}^{\mathcal{B}} (\epsilon) \approx E_{\text{FCI}}^{\mathcal{B}} (\epsilon) 
 + \bar{E}^{\mathcal{B}}[n_{\text{FCI}}^{\basis} (\epsilon)] ,
\end{equation}
where $n_{\text{FCI}}^{\basis}(\epsilon)$ is the ground state FCI density obtained in the presence of the electric field of strength $\epsilon$. The corresponding non-self-consistent basis-set corrected dipole moment is thus
\begin{equation}
 \begin{aligned}
	\label{eq:dipole_form}
	d^{\mathcal{B}} & \approx d^{\mathcal{B}}_{\text{FCI}} + \bar{d}^{\mathcal{B}},
 \end{aligned}
\end{equation}
where 
\begin{equation}
 \label{eq:d_bar_b}
  \bar{d}^{\mathcal{B}} = - \deriv{\bar{E}^{\mathcal{B}}[n_{\text{FCI}}^{\basis} (\epsilon)]}{\epsilon}{}  
\end{equation}
is the non-self-consistent basis-set correction to the dipole moment. 

As obtaining both the dipole moment and the density at FCI level is often computationally prohibitive, we follow Ref.~\onlinecite{LooPraSceTouGin-JPCL-19} and approximate the FCI energy by the CCSD(T) energy and the FCI density by the HF density (in the presence of the electric field)
\begin{equation}
	\label{eq:REFG2}
	E_{0}^{\mathcal{B}} (\epsilon) \approx E_{\text{CCSD(T)}}^{\mathcal{B}} (\epsilon) + \bar{E}^{\mathcal{B}}[n_{\text{HF}}^{\basis} (\epsilon)].
\end{equation}
Within these approximations, the basis-set corrected dipole moment in Eq.~\eqref{eq:dipole_form} becomes
\begin{equation}
	\label{eq:NONSCFDipole}
	\begin{split}
		d^{\mathcal{B}} &\approx d_{\text{CCSD(T)}}^{\mathcal{B}} + \bar{d}^{\mathcal{B}},
	\end{split}
\end{equation}
where $d_{\text{CCSD(T)}}^{\mathcal{B}}$ is the dipole moment at the CCSD(T) level and the basis-set correction $\bar{d}^{\mathcal{B}}$ is
\begin{equation}
 \bar{d}^{\mathcal{B}} = - \deriv{\bar{E}^{\mathcal{B}}[n_{\text{HF}}^{\basis} (\epsilon)]}{\epsilon}{}.
\end{equation}

We approximate the basis-set correction functional $\bar{E}^{\mathcal{B}}[n]$ with the so-called (spin-dependent) PBEUEG energy functional introduced in Ref.~\onlinecite{LooPraSceTouGin-JPCL-19} where the local range-separation parameter $\mu^{\mathcal{B}}(\bm{r})$ is obtained using the HF wave function in the basis set $\basis$ as proposed in Refs. \onlinecite{GinPraFerAssSavTou-JCP-18} and \onlinecite{LooPraSceTouGin-JPCL-19}. The results obtained with Eq.~\eqref{eq:NONSCFDipole} with the PBEUEG approximation for $\bar{E}^{\mathcal{B}}[n]$ evaluated at the HF density will be referred to as CCSD(T)+PBEUEG.

In practice, we calculate the CCSD(T) dipole moment and the basis-set correction to the dipole moment using a finite-difference approximation for the energy derivatives with respect to the electric field
\begin{equation}
	d_{\text{CCSD(T)}}^{\mathcal{B}} \simeq - \frac{E_{\text{CCSD(T)}}^{\mathcal{B}} \left( \epsilon \right) - E_{\text{CCSD(T)}}^{\mathcal{B}} \left( - \epsilon \right) }{2 \epsilon }, 
\end{equation}
and 
\begin{equation}
	\bar{d}^{\mathcal{B}} \simeq - \frac{\bar{E}^{\mathcal{B}} \left[  n_{\text{HF}}^\basis(\epsilon) \right] - \bar{E}^{\mathcal{B}} \left[  n_{\text{HF}}^\basis(-\epsilon) \right] }{2 \epsilon},
\end{equation}
using a finite field strength of $\epsilon = 10^{-4}$ a.u., as suggested in Ref.~\onlinecite{HalKloHelJor-JCP-99}.

\section{Computational details}\label{sec:Comput}

The computation of the basis-set correction to the dipole moment $\bar{d}^{\mathcal{B}}$ were performed using the Quantum Package program \cite{QP2} and the CCSD(T) dipole moment were obtained with the Gaussian program \cite{g16}. 
We used the augmented Dunning basis sets (Refs.~\onlinecite{KenDunHar-JCP-92,PraWooPetDunWil-TCA-2011}) aug-cc-pVXZ (abbreviated as AV$X$Z in the tables and figures of the paper) where $X$ is the cardinal number of the basis set $\text{X}\in\{\text{D}, \text{T}, \text{Q}, \text{5}\}$. 
As no core-valence functions are used, the frozen-core approximation is used throughout this paper where the 1s orbital is kept frozen for the elements from Li to F.

The tests are done on a set of $n=14$ molecules among which six open-shell molecules, for which we use restricted open-shell CCSD(T) (ROCCSD(T)) energies and restricted open-shell HF (ROHF) densities, and eight closed-shell molecules. Experimental geometries used for the computations are taken from Ref.~\onlinecite{HaiHea-JCTC-18} for the entire set except in the case of BH and FH for which the geometries are taken from Ref.~\onlinecite{HalKloHelJor-JCP-99}. 
We also report the results obtained in Ref.~\onlinecite{GinTraPraTou-JCP-21} for the 
BH, FH, CH$_2$, and H$_2$O molecules using the self-consistent formalism [Eq.~\eqref{eq:SCFDipole}] at near-FCI level in order to compare with the present non-self-consistent formalism. 

The accuracy of the dipole moments obtained with a given basis set and a given level of approximation is evaluated 
with respect to the CBS limit of the CCSD(T) dipole moments, $d^{\text{CBS}}_\text{CCSD(T)}$, 
which are evaluated as in Ref.~\onlinecite{HalKloHelJor-JCP-99}. Namely, the CBS results are computed as follows 
\begin{equation}\label{eq:CBSDipole}
	d^{\text{CBS}}_\text{CCSD(T)} = d^{\text{CBS}}_{\text{HF}}  + d^{\text{CBS}}_{\text{c}},
\end{equation}
where $d^{\text{CBS}}_{\text{c}}$ is the CBS limit of the correlation contribution to the CCSD(T) dipole moment which is computed using the following two-point $X^{-3}$ extrapolation formula
\begin{equation}\label{eq:HelExtrap}
	d^{\text{CBS}}_{\text{c}} = \frac{d^{\text{X}}_{\text{c}} X^3 - d^{\text{(X-1)}}_{\text{c}} (X-1)^3}{ X^{3} - (X-1)^{3}},
\end{equation}
with 
\begin{equation}
	d_{\text{c}}^{\text{X}} = d^{\text{X}}_{\text{CCSD(T)}} - d_{\text{HF}}^{\text{X}},
\end{equation}
where $d_{\text{c}}^{\text{X}}$ and $d_{\text{HF}}^{\text{X}}$ are the correlation and HF contributions, respectively, to the CCSD(T) dipole moment using the aug-cc-pV$X$Z basis set. 
In the present calculations, we evaluate Eq.~\eqref{eq:HelExtrap} at $X=5$ and we estimate the CBS limit of HF dipole moment
$d^{\text{CBS}}_{\text{HF}}$ in Eq.~\eqref{eq:CBSDipole} simply by using the HF dipole moment in the aug-cc-pV$5$Z basis set. 
For all the systems studied here, the HF dipole moments are converged with an accuracy better than 0.001 a.u. (as measured by the difference between the aug-cc-pVQZ and aug-cc-pV5Z dipole moments).

At a given level of calculation in a basis set $\basis$ we report the error on the dipole moment with respect to the CBS reference $\Delta^{\mathcal{B}}=d^{\mathcal{B}} - d^{\text{CBS}}_\text{CCSD(T)}$ and the relative error $\Delta^{\mathcal{B}}_{\text{rel}} = \Delta^{\mathcal{B}}/d^{\text{CBS}}_\text{CCSD(T)}$.
To statistically analyse the results, we also calculate the normal distribution function of the errors for a given basis set $\mathcal{B}$,
\begin{equation}
	\rho (\Delta^{\mathcal{B}}) = \frac{1}{\Delta_{\text{std}}^{\mathcal{B}} \sqrt{2\pi}} \text{exp} \left[ -\frac{1}{2} \left( \frac{\Delta^{\mathcal{B}} - \bar{\Delta}^{\mathcal{B}}}{\Delta_{\text{std}}^{\mathcal{B}}} \right)^2 \right],
	\label{Eq:NormalDistri}
\end{equation}
where $\bar{\Delta}^{\mathcal{B}} =  (\sum_{i=1}^{n} \Delta^{\mathcal{B}}_{i})/n$ is the mean error and $\Delta_{\text{std}}^{\mathcal{B}} = \sqrt{\sum_{i=1}^{n} (\Delta^{\mathcal{B}}_{i} - \bar{\Delta}^{\mathcal{B}})^2/(n-1)}$ is the root-mean-square deviation.

\section{Results and discussion}\label{sec:Results}

\begin{figure*}[!htb]
 \centering
       	\includegraphics[width=0.47\linewidth]{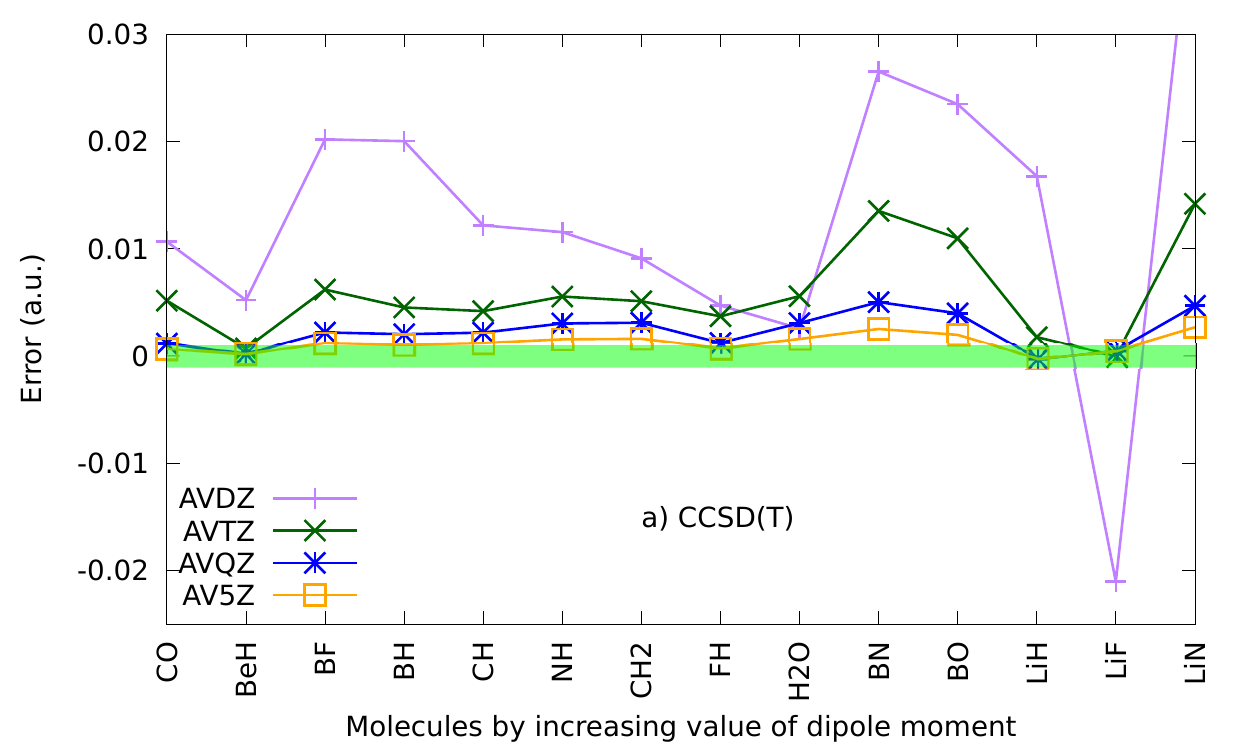}
       	\includegraphics[width=0.47\linewidth]{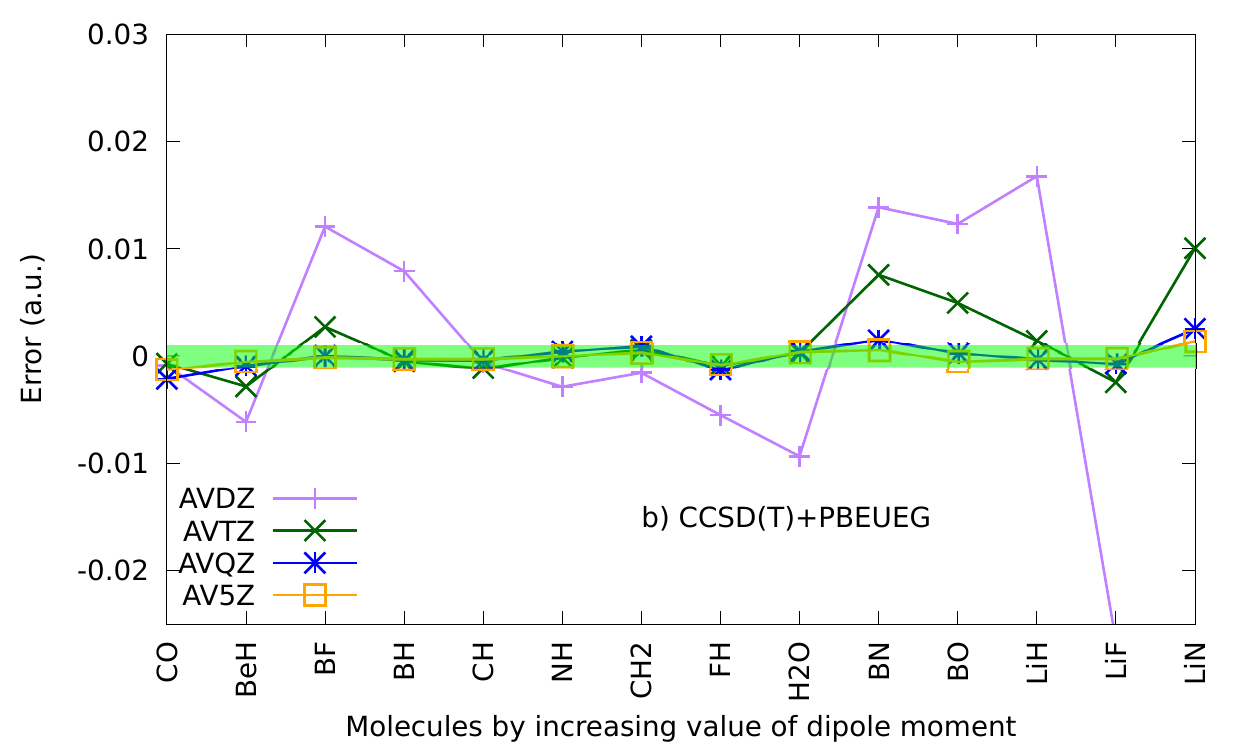}
	\captionof{figure}{\label{fig:error_ccsdt_new_sel}(a) CCSD(T) and (b) CCSD(T)+PBEUEG errors on the dipole moments of 14 molecules compared to CCSD(T)/CBS reference values. The green area indicates an error of $\pm$0.001 a.u..}
\end{figure*}

 \begin{figure*}[!htb]
 \centering
        	\includegraphics[width=0.47\linewidth]{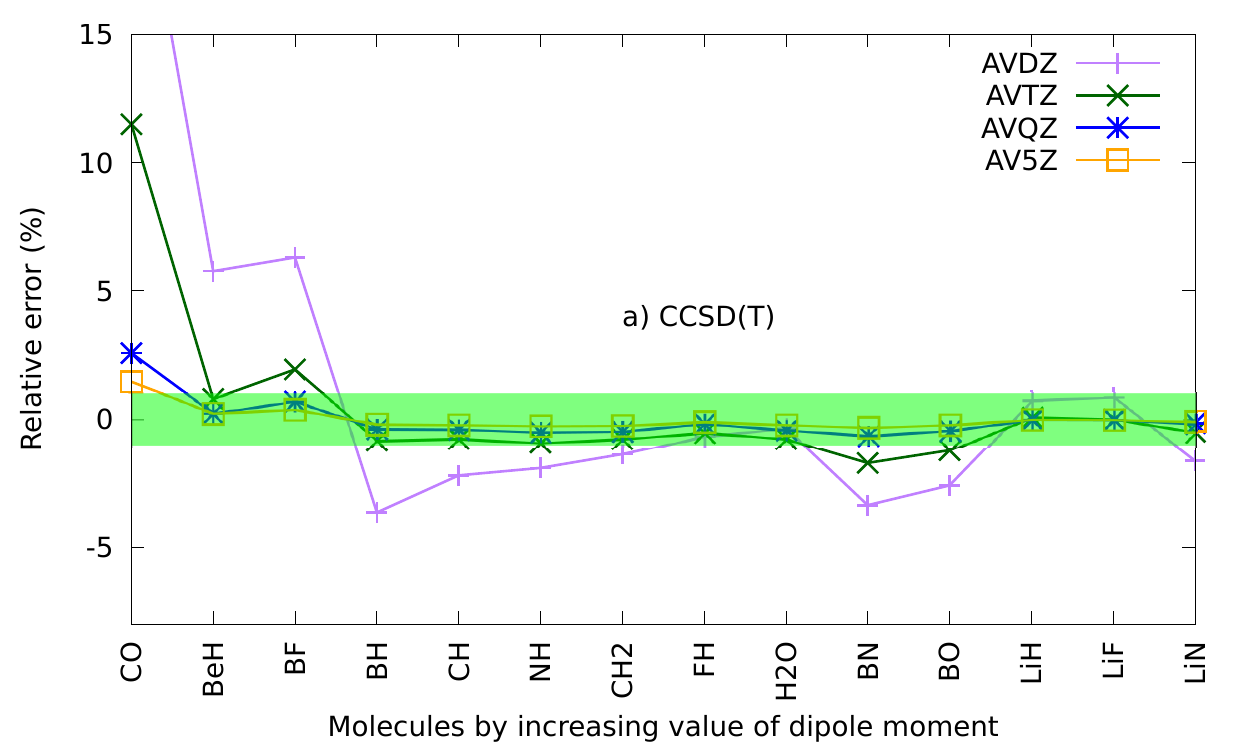}
        	\includegraphics[width=0.47\linewidth]{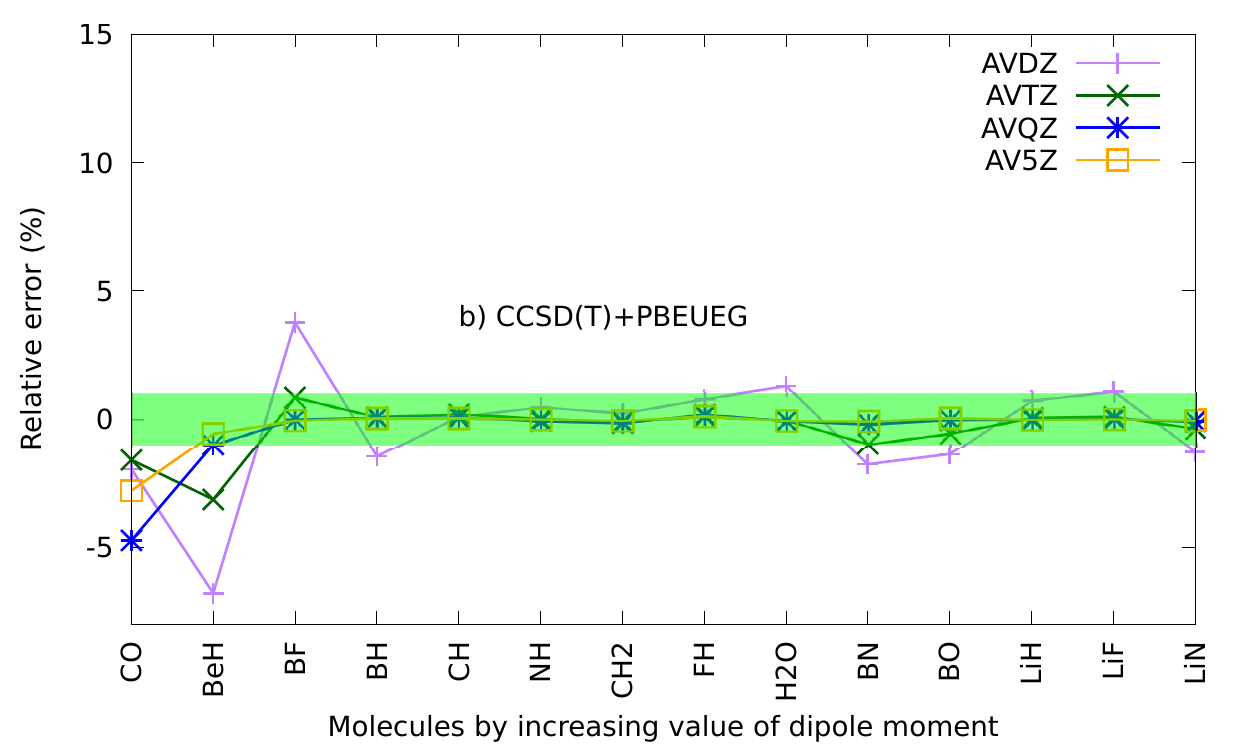}
	\captionof{figure}{\label{fig:rel_error_ccsdt_new_sel}(a) CCSD(T) and (b) CCSD(T)+PBEUEG relative errors on the dipole moments of 14 molecules compared to CCSD(T)/CBS reference values. The green area indicates an error of $\pm$1\%.}
 \end{figure*}

 \begin{figure*}[!htb]
 \centering
 		\includegraphics[width=0.47\linewidth]{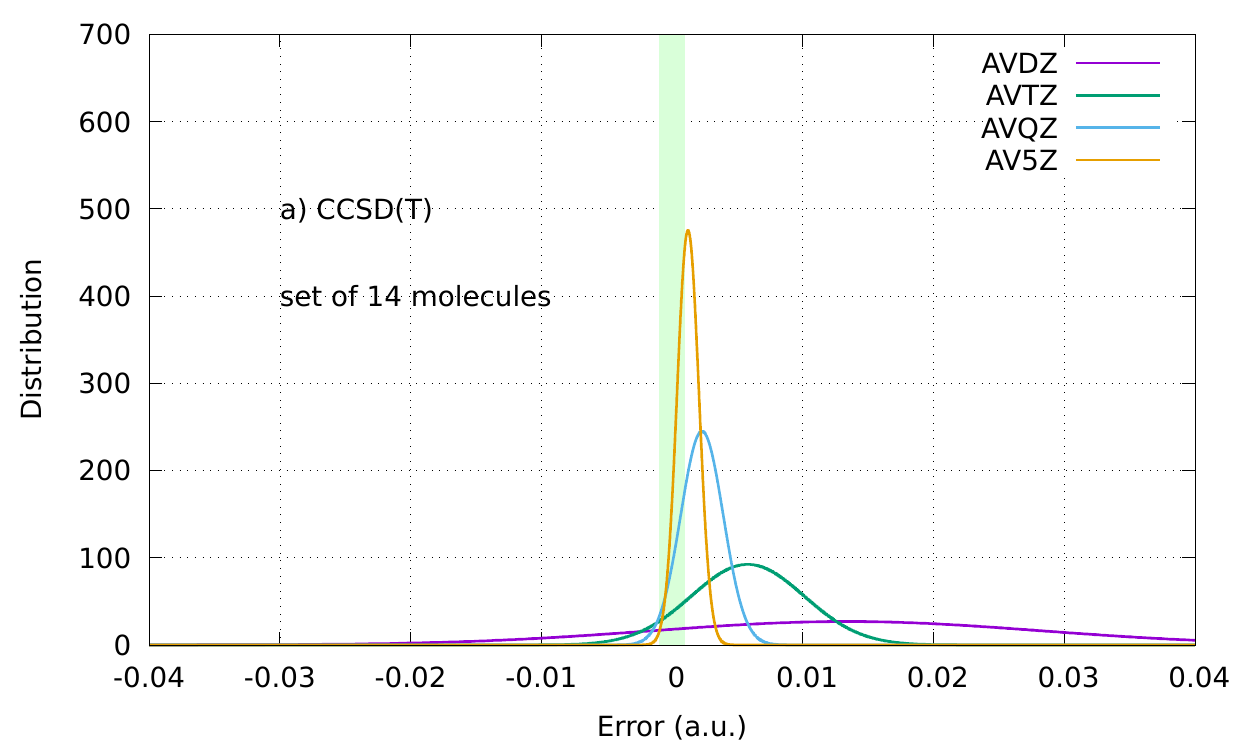}
 		\includegraphics[width=0.47\linewidth]{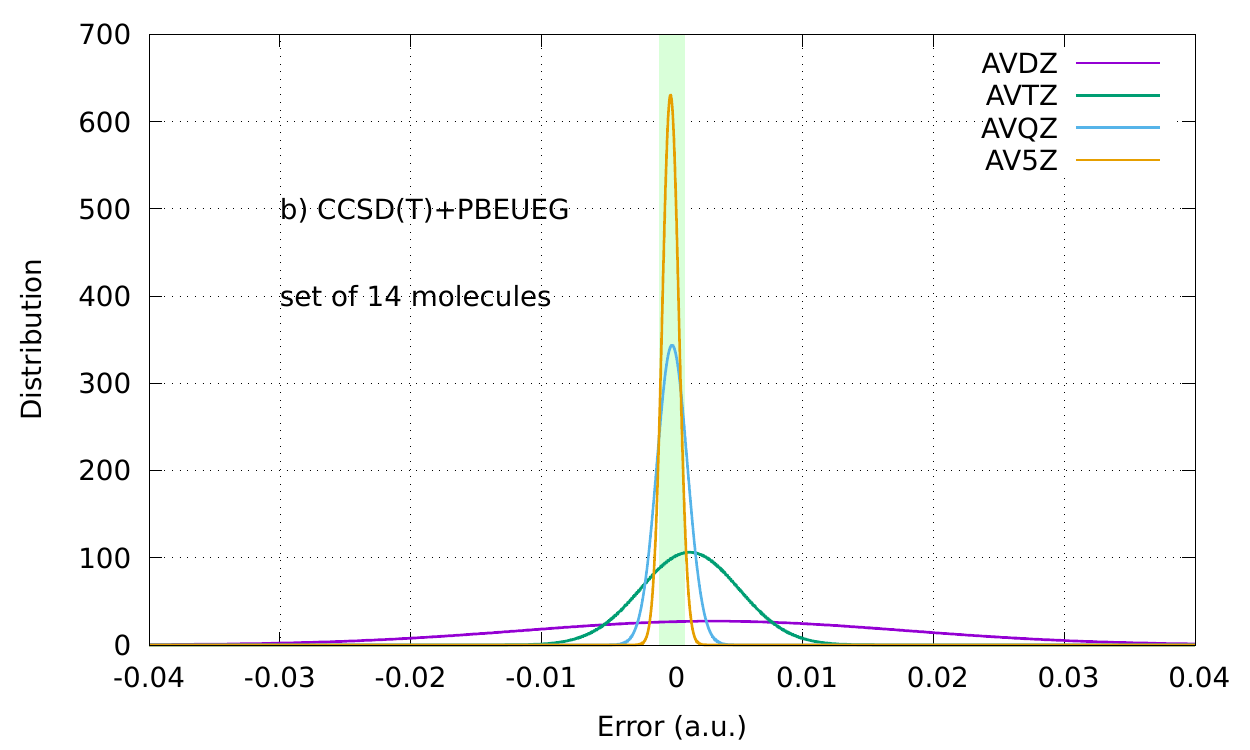}
	\captionof{figure}{\label{plot_error_cc_closed_good_hf}(a) CCSD(T) and (b) CCSD(T)+PBEUEG normal distribution of errors on the dipole moments of 14 molecules compared to CCSD(T)/CBS reference values. The green area indicates an error of $\pm$ 0.001 a.u..}
 \end{figure*}

In Table~\ref{Tab:14molecules}, we report the dipole moments at various levels of approximations (HF, CCSD(T), CCSD(T)+PBEUEG) with different basis sets, as well as the CCSD(T)/CBS reference values, for the set of 14 molecules. Note the wide range of magnitudes of the dipole moments (from 0.04485 a.u. for CO to 2.78718 a.u. for LiN).
The mean error (ME), mean absolute error (MAE), mean absolute relative error (MARE), maximal absolute error (MAX), and root-mean-square deviation (RMSD) obtained with CCSD(T) and CCSD(T)+PBEUEG are reported in Table \ref{Tab:NormalDistr_selected}. 
The graphical representations of this data are provided in Figs.~\ref{fig:error_ccsdt_new_sel} and~\ref{fig:rel_error_ccsdt_new_sel} for the errors and relative errors, and in Fig.~\ref{plot_error_cc_closed_good_hf} for the normal distributions of errors.

Analyzing first the results at the CCSD(T) level in Table \ref{Tab:NormalDistr_selected}, one can notice that, 
 as expected, the ME and MAE systematically decrease with the size of the basis set. 
Moreover, as noticed in previous studies \cite{TewKloNeiHat-PCCP-2017,LooPraSceTouGin-JPCL-19}, 
not only the average values of the errors but also the RMSD tends to decrease 
with the basis-set size.
Nevertheless, the improvement of the results is rather slow as a MAE below 0.001 a.u. is not reached even with the aug-cc-pV5Z basis set, illustrating the slow convergence of properties with respect to the basis set at the CCSD(T) level.    
Regarding the relative errors in Fig.~\ref{fig:rel_error_ccsdt_new_sel}, one can notice that the largest errors with respect to the CBS reference come from the molecules with smallest dipole moments (\textit{i.e.} CO and BeH). 
More quantitatively, an aug-cc-pVQZ basis set is needed to obtain a MARE smaller than 1$\%$. 

Going from CCSD(T) to CCSD(T)+PBEUEG, one observes a systematic decrease of the MAE, ME, MARE, and RMSD.  
Focusing on the MAE, an error below 0.001 a.u. is reached with the aug-cc-pVQZ basis set, whereas such an accuracy is not even reached at the CCSD(T) level with the aug-cc-pV5Z basis set. 
Qualitatively, for the aug-cc-pVTZ basis set and larger basis sets, the MAEs obtained with CCSD(T)+PBEUEG with a basis set of cardinal number $X$ are comparable to the MAEs obtained with CCSD(T) with a basis set of cardinal number $X$+1. 
Regarding the MARE, an error below 1$\%$ is reached with CCSD(T)+PBEUEG already with the aug-cc-pVTZ basis set.
One nevertheless observes that the effects of the basis-set correction on the RMSD is very weak. 
From the plots of Fig.~\ref{fig:error_ccsdt_new_sel} one notices that, even if the basis-set correction 
systematically improves the results for the aug-cc-pVTZ basis set, its effect is less impressive 
when there is both a large error and a large dipole moment (\textit{i.e.} for BN, BO, and LiN). 

We conclude this study by confronting the present CCSD(T)+PBEUEG results with the results obtained in Ref.~\onlinecite{GinTraPraTou-JCP-21} with the self-consistent basis-set correction formalism using near-FCI (CIPSI) wave functions for 4 molecules which are reported in Table~\ref{table_scf}. One notices that the non-self-consistent CCSD(T)+PBEUEG results are in good agreement with the self-consistent  CISPI+PBEUEG results.
More quantitatively, the absolute deviation between CCSD(T)+PBEUEG and self-consistent CIPSI+PBEUEG in a given basis set is never larger than 0.001 a.u. for FH and H$_2$O, and the discrepancy slightly increases up to 0.006 and 0.003 a.u. in the case of CH$_2$ and BH, respectively. Nevertheless, as originally reported in Ref.~\onlinecite{GinTraPraTou-JCP-21} and apparent from Table~\ref{table_scf}, discrepancies of the same order of magnitude also appear between the uncorrected CIPSI and CCSD(T) results in the case of the CH$_2$ and BH molecules.
This suggests that the main source of difference between the CCSD(T)+PBEUEG and self-consistent CIPSI+PBEUEG methods
actually comes from the parent WFT theory.
These results illustrate the validity of the different approximations leading to the CCSD(T)+PBEUEG approach and are encouraging considering that the latter has a much lower computational cost with respect to the self-consistent basis-set formalism. Indeed, CCSD(T)+PBEUEG relies only on a standard CCSD(T) calculation and a HF calculation for the basis-set correction which is of negligible computational cost with respect to CCSD(T). 

 \begin{table*}[!htb]
       \caption{\label{Tab:14molecules}HF, CCSD(T), and CCSD(T)+PBEUEG dipole moments in atomic units. For the open-shell systems, we use the spin-restricted open-shell (RO) version of these methods.}
 \begin{ruledtabular}
 \begin{tabular}{llllll}
        &      \multicolumn{1}{c}{AVDZ}   &         \multicolumn{1}{c}{AVTZ}     &  \multicolumn{1}{c}{AVQZ}     & \multicolumn{1}{c}{AV5Z} & \multicolumn{1}{c}{CBS}             \\
	\hline
	 \textbf{CO} & & & & &  \\
        HF & -0.10199 & -0.10499 & -0.10433 & -0.10421 & \\ 
        CCSD(T) & 0.05550 & 0.05000 & 0.04600 & 0.04550 & 0.04485 \\
	CCSD(T)  + PBEUEG & 0.04398 & 0.04414 & 0.04273 & 0.04360 & \\ [0.2cm]
	 \textbf{BeH} & & & & & \\
        ROHF & 0.11017 & 0.11076 & 0.11199 & 0.11218 & \\
        ROCCSD(T) & 0.09550 & 0.09100 & 0.09050 & 0.09050 & 0.09030 \\
	 ROCCSD(T)  + PBEUEG & 0.08416 & 0.08746 & 0.08941 & 0.08980 & \\ [0.2cm]
	 \textbf{BF} & & & & & \\
        HF & 0.34436 & 0.33390 & 0.33314 & 0.33328 & \\
        CCSD(T) & 0.34100 & 0.32700 & 0.32300 & 0.32200 & 0.32081 \\
	 CCSD(T)  + PBEUEG & 0.33287 & 0.32351 & 0.32082 & 0.32068 & \\ [0.2cm]
	 \textbf{BH} & & & & & \\ 
        HF & 0.68796 & 0.68649 & 0.68494 & 0.68496 & \\ 
        CCSD(T)           & 0.52950 & 0.54500 & 0.54750 & 0.54850 & 0.54953 \\ 
	 CCSD(T)+PBEUEG & 0.54162 & 0.55002 & 0.54986 & 0.54980 & \\ [0.2cm]
	 \textbf{CH} & & & & & \\
        ROHF & 0.62348 & 0.62000 & 0.61871 & 0.61858 &  \\
        ROCCSD(T)                 & 0.54150 & 0.54950 & 0.55150 & 0.55250 & 0.55368\\
	 ROCCSD(T)  + PBEUEG    & 0.55427 & 0.55481 & 0.55405 & 0.55396 & \\ [0.2cm]
	 \textbf{NH} & & & & & \\
        ROHF & 0.63850 & 0.63505 & 0.63381 & 0.63384 & \\ 
        ROCCSD(T) & 0.59350 & 0.59950 & 0.60200 & 0.60350 & 0.60504 \\
	 ROCCSD(T)  + PBEUEG & 0.60792 & 0.60519 & 0.60464 & 0.60506 & \\[0.2cm]
	 \textbf{CH$_2$ (singlet)} & & & & & \\
        HF & 0.74877 & 0.74477 & 0.74355 & 0.74353 & \\ 
        CCSD(T) & 0.65600 & 0.66000 & 0.66200 & 0.66350 &  0.66510 \\
	 CCSD(T)  + PBEUEG & 0.66666 & 0.66455 & 0.66420 & 0.66478 & \\ [0.2cm] 
	 \textbf{FH} & & & & & \\
         HF & 0.75976 & 0.75751 & 0.75634 & 0.75617 & \\
         CCSD(T) & 0.70350 & 0.70450 & 0.70700 & 0.70750 & 0.70820 \\
	 CCSD(T)  + PBEUEG & 0.71371 & 0.70903 & 0.70946 & 0.70900 & \\ [0.2cm]
	 \textbf{H$_2$O} & & & & & \\
         HF & 0.78671 & 0.78039 & 0.77956 & 0.77956 & \\ 
         CCSD(T) & 0.72700 & 0.72400 & 0.72650 & 0.72800 & 0.72957 \\
	 CCSD(T)  + PBEUEG & 0.73891 & 0.72930 & 0.72912 & 0.72920 & \\ [0.2cm]
         \textbf{BN} & & & & & \\
         ROHF & 1.13451 & 1.13862 & 1.13831 & 1.13840 & \\ 
         ROCCSD(T) & 0.76250 & 0.77550 & 0.78400 & 0.78650 & 0.78902 \\
	 ROCCSD(T)  + PBEUEG & 0.77517 & 0.78145 & 0.78756 & 0.78846 & \\ [0.2cm]
	 \textbf{BO} & & & & & \\
         ROHF & 1.17803 & 1.18533 & 1.18527 & 1.18539 & \\ 
         ROCCSD(T) & 0.88300 & 0.89550 & 0.90250 & 0.90450 & 0.90647 \\
	 ROCCSD(T)  + PBEUEG & 0.89417 & 0.90153 & 0.90622 & 0.90698 & \\ [0.2cm]
	 \textbf{LiH} & & & & & \\ 
         HF & 2.37055 & 2.36235 & 2.36153 & 2.36129 & \\
         CCSD(T) & 2.32500 & 2.31000 & 2.30800 & 2.30800 & 2.30825 \\ 
	 CCSD(T)+PBEUEG & 2.32501 & 2.30965 & 2.30795 & 2.30802 & \\ [0.2cm]
	 \textbf{LiF} & & & & & \\
         HF & 2.56111 & 2.54103 & 2.53949 & 2.53905 & \\ 
         CCSD(T)                  & 2.50400 & 2.48300 & 2.48250 & 2.48250 & 2.48297 \\
	 CCSD(T)  + PBEUEG    & 2.50942 & 2.48542 & 2.48367 & 2.48321 & \\ [0.2cm]
	 \textbf{LiN} \\ 
         ROHF & 2.90309 & 2.90379 & 2.90372 & 2.90317 & \\
         ROCCSD(T) & 2.74200 & 2.77300 & 2.78250 & 2.78450 & 2.78718 \\ 
	 ROCCSD(T)  + PBEUEG & 2.75215 & 2.77714 & 2.78464 & 2.78583 \\ [0.2cm]
 \end{tabular}
 \end{ruledtabular}
 \end{table*}

\begin{table*}[!htb]
	\captionof{table}{\label{Tab:NormalDistr_selected}Mean error (ME), mean absolute error (MAE), mean absolute relative error (MARE), maximal absolute error (MAX), and root-mean-square deviation (RMSD) (in atomic units) for the CCSD(T) and CCSD(T)+PBEUEG dipole moments of 14 molecules. See Fig.~\ref{plot_error_cc_closed_good_hf} for the corresponding plots of the normal distributions of errors.}
 \begin{ruledtabular}
 \begin{tabular}{lllll}
         & AVDZ & AVTZ & AVQZ & AV5Z \\
         \hline
	 \textbf{ME}           & & & & \\
         CCSD(T)               & 0.01336 & 0.00579 & 0.00229  &  0.00122 \\
	 CCSD(T)+PBEUEG        & 0.00319 & 0.00135 & 0.000004 & -0.00012 \\ [0.2cm]
	 \textbf{MAE}          & & & & \\
         CCSD(T)               & 0.01637 & 0.00579 & 0.00233 & 0.00125 \\
	 CCSD(T)+PBEUEG        & 0.01080 & 0.00258 & 0.00086 & 0.00049 \\ [0.2cm]
	 \textbf{MARE (in \%)} & & & & \\
         CCSD(T)               & 3.9 & 1.5 & 0.5 & 0.3 \\ 
	 CCSD(T)+PBEUEG        & 1.6 & 0.6 & 0.4 & 0.3 \\ [0.2cm] 
	 \textbf{MAX}          & & & & \\
	 CCSD(T)               & 0.04518 (LiN) & 0.01418 (LiN) & 0.00502 (BN)  & 0.00268 (LiN) \\
	 CCSD(T)+PBEUEG        & 0.03504 (LiN) & 0.01004 (LiN) & 0.00254 (LiN) & 0.00136 (LiN) \\ [0.2cm]
	 \textbf{RMSD}         & & & & \\
	 CCSD(T)               & 0.01484 & 0.00432 & 0.00163 & 0.00084 \\ 
	 CCSD(T)+PBEUEG        & 0.01464 & 0.00376 & 0.00116 & 0.00063 \\ [0.2cm]
 \end{tabular}
 \end{ruledtabular}
 \end{table*}

\section{Conclusion}\label{sec:Conclu}

In the present study, we have proposed an extension of the recently introduced non-self-consistent basis-set 
correction of CCSD(T) ground-state energies\cite{LooPraSceTouGin-JPCL-19} to the computation of properties as energy derivatives, focussing here on the dipole moment. 
The theory relies on the originally proposed DFT-based basis-set correction approach\cite{GinPraFerAssSavTou-JCP-18} 
which accelerates the basis-set convergence to the unaltered CBS limit.  
Numerical tests on a set of 14 molecules (including both closed and open-shell) with dipole moments spanning two orders of magnitude have been carried in order to obtain a representative study of the performance of the present approach. 

Although this study aims at correcting the basis-set convergence of the CCSD(T) dipole moments, it can be formally generalized 
to any wave-function method and any energy derivative with respect to a static perturbation.
In its present form, the basis-set correction relies only on HF calculations, which makes the basis-set correction essentially computationally free compared to the correlated wave-function calculation.
This approach is an alternative to the recently proposed self-consistent basis-set 
correction\cite{GinTraPraTou-JCP-21} which allows for the computation of first-order properties through 
expectation values over an energy-minimized wave function. 
In contrast with the self-consistent formalism, the present approach does not require a variational wave function, 
which considerably extend the domain of application of the basis-set correction. 

Regarding now the numerical results, we have shown that the present approach significantly accelerates the basis-set convergence of CCSD(T) dipole moments.
Typically, the error obtained in a basis set of cardinal $X$ with the basis-set correction is comparable to the error of the uncorrected CCSD(T) calculation with cardinal number $X$+1.
We also compared the present non-self-consistent basis-set correction with the self-consistent formalism of Ref.~\onlinecite{GinTraPraTou-JCP-21} 
and shown that the two theories agree within a few milli-atomic units, illustrating the soundness of the approximations leading to the non-self-consistent approach. 

Considering the generality, the global performance, and the small computational cost of the present approach, it could be an alternative to explicitly correlated approaches for calculation of molecular properties. In the near future we will extend the method to higher-order static properties, such as static polarizabilities, and also to more general dynamic properties, leading in particular to the possibility of accelerating the basis-set convergence of excitation energies.

\section*{Acknowledgement}
This project has received funding from the European Research Council (ERC) under the European Union's Horizon 2020 research and innovation programme Grant agreement No. 810367 (EMC2).

\section*{Author Declarations}
The authors have no conflicts to disclose.

\section*{Data Availability}
The data presented in this study are available in the supplementary materials associated with the present paper. 

 \begin{table*}[!htb]
       \caption{\label{table_scf} Dipole moments obtained with near FCI (CIPSI) calculations and self-consistent basis-set correction (SC CIPSI+PBEUEG) from Ref.~\onlinecite{GinTraPraTou-JCP-21}, compared to the present CCSD(T) and CCSD(T)+PBEUEG results. Estimated CBS values using Eq.~\eqref{eq:HelExtrap} with $X=5$ are reported when computations could be done with the aug-cc-pV5Z basis set.}
 \begin{ruledtabular}
 \begin{tabular}{llllll}
        &      \multicolumn{1}{c}{AVDZ}   &         \multicolumn{1}{c}{AVTZ}     &  \multicolumn{1}{c}{AVQZ}     & \multicolumn{1}{c}{AV5Z} & \multicolumn{1}{c}{CBS}             \\
	\hline
         \textbf{BH} & & & & & \\ 
	 CIPSI$^a$ & 0.52782 & 0.54334 & 0.54563 & 0.54691 & 0.54823 \\
         CCSD(T)           & 0.52950 & 0.54500 & 0.54750 & 0.54850 & 0.54953 \\ 
	 SC CIPSI+PBEUEG$^a$ & 0.53791 & 0.54815 & 0.54790 & 0.54815 &  \\ 
	 CCSD(T)+PBEUEG & 0.54162 & 0.55002 & 0.54986 & 0.54980 & \\ [0.2cm]
	 \textbf{CH$_2$ (singlet)} & & & & & \\
	 CIPSI$^a$ & 0.65120 & 0.65446 & 0.65643 & 0.65780 & 0.65926 \\
         CCSD(T) & 0.65600 & 0.66000 & 0.66200 & 0.66350 & 0.66510 \\
	 SC CIPSI+PBEUEG$^a$ & 0.66249 & 0.65958 & 0.65890 & ---$^b$ &  \\ 
	 CCSD(T)+PBEUEG & 0.66666 & 0.66455 & 0.66420 & 0.66478 & \\ [0.2cm] 
	 \textbf{FH} & & & & & \\
	 CIPSI$^a$ & 0.70249 & 0.70406 & 0.70662 & ---$^b$ & ---$^b$    \\
         CCSD(T) & 0.70350 & 0.70450 & 0.70700 & 0.70750 & 0.70820 \\
	 SC CIPSI+PBEUEG$^a$ & 0.71326 & 0.70873 & ---$^b$ & ---$^b$ &  \\ 
	 CCSD(T)+PBEUEG & 0.71371 & 0.70903 & 0.70946 & 0.70900 & \\ [0.2cm]
	 \textbf{H$_2$O} & & & & & \\
	 CIPSI$^a$ & 0.72610 & 0.72294 & ---$^b$ & ---$^b$ & ---$^b$\\
         CCSD(T) & 0.72700 & 0.72400 & 0.72650 & 0.72800 & 0.72957 \\
	 SC CIPSI+PBEUEG$^a$ & 0.73809 & 0.72818 & ---$^b$ & ---$^b$& \\ 
	 CCSD(T)+PBEUEG & 0.73891 & 0.72930 & 0.72912 & 0.72920 & \\ [0.2cm]
      \end{tabular}
 \end{ruledtabular}
$^a$ From Ref.~\onlinecite{GinTraPraTou-JCP-21}.\\
$^b$ Results non available due to the computational requirement.
 \end{table*}


\end{document}